\documentclass[fleqn,usenatbib]{mnras}

\usepackage[T1]{fontenc}

\usepackage{graphicx}	
\usepackage{amsmath}	
\usepackage{amssymb}	

\title[  power spectrum of the solar surface magnetic turbulence]{Interpretation of the power spectrum of the  quiet Sun photospheric turbulence}  
\author[I.Goldman]{Itzhak Goldman$^{1, 2}$  \thanks{E-mail:goldman@afeka.ac.il} \\  
$^1$ Department of Physics, Afeka College, Tel Aviv, Israel \\
$^2$ Department of Astrophysics, Tel Aviv University, Tel Aviv, Israel }
 \pubyear{2020}

\begin{document}
\label{firstpage}
\pagerange{\pageref{firstpage}--\pageref{lastpage}}
\maketitle 

\begin{abstract} 
 Observational  power  spectra of the   photospheric  magnetic field turbulence, of the quiet-sun, were presented in a recent paper by Abramenko \& Yurchyshyn. Here I focus on the power spectrum   derived from the observations of the Near InfraRed
Imaging Spectrapolarimeter (NIRIS) operating at the Goode Solar Telescope.  The latter  exhibits  a transition from a power law with index $-1.2$  to a steeper power law with index $-2.2$, for smaller spatial scales. The present paper presents an interpretation of this  change. Furthermore, this interpretation  provides an estimate for the  effective width of the turbulent  layer probed by the observations.  The latter turns out to be  practically equal to the depth of the photosphere. 
\end{abstract}

\begin{keywords}
turbulence-solar photosphere- magnetic field
\end{keywords} 
 
\section{Introduction} 
  
  Quite recently,    \cite{Abramenko+Yurchyshyn2020}  measured the line-of-sight magnetic fields and derived magnetic power spectra of the quiet sun photosphere. The data was obtained from  magnetograms of the Helioseismic and Magnetic Imager (HMI) on-board the Solar Dynamic Observatory (SDO), and the Near  Infrared
Imaging Spectrapolarimeter (NIRIS) operating at the Goode Solar Telescope (GST) of the Big Bear Solar Observatory(BBSO).  
 The authors computed 1-dimensional power spectra of the observed line-of-sight turbulent magnetic field. 
For data from both observatories, the power spectra inertial range  were considerably shallower than the Kolmogorov spectrum characterized by an index  of $-5/3$.(\cite{Kolmogorov1941}).

An interesting feature is revealed in the NIRIS power spectrum (\cite{Abramenko+Yurchyshyn2020}, Fig. 6) derived from measurements 
using the Fe I 15650Å. While   the inertial range on spatial scales larger than 
($0.8\div 0.9$)~Mm has a logarithmic slope of $-1.2$, the logarithmic slope on smaller scales is $-2.2$. The NIRIS data extends down to  spatial scale of $\sim 0.3$~Mm.  The above transition  is not present in the HMI data, as the HMI resolution is about 2.4~Mm.

  The magnetometer  data originate from an integral along the line-of-sight of the emitted near infra-red  radiation. The Zeeman splitting of emission lines  is used to  obtain the line-of-sight magnetic field. For more details on such measurements see  \cite{Abramenko+2001} and references therein.
The focus of the present paper is on   the two logarithmic slopes in the power spectrum of the NIRIS data. 

It is worth noting that 
 several authors addressed the issue of power spectra of quantities which  are the result of integration along the line-of-sight (see  e.g. \cite{Stutzki+98}, \cite{Goldman2000},   \cite{Lazarian+Pogosyan2000},  \cite{Miville+apj2003}). They concluded that when the lateral spatial scale is much smaller than the depth of the layer, the logarithmic slope steepens exactly by $-1$ compared to its value when the lateral scale is much larger than the depth. 
 
 This behavior was indeed  found in observational power spectra  of Galactic and extra Galactic turbulence (e.g. 
\cite{Elmegreen+2001}, \cite{Miville+2003}, \cite{Block+2010} and  \cite{contini+goldman2011}).

Here,  the same turbulence phenomenon is addressed in the context of the solar photosphere, where the measured  magnetic field   follows from the sum of the infra-red radiation emanating from various depths. 

In   section 2  the theoretical 1-D power spectrum of a, line-of-sight integrated data, is explicitly derived.
 
In section 3 the asymptotic behavior of the power spectrum in the small and large wavenumber limits is obtained analytically. In section 4., the numerical results for the specific case  under consideration here, are obtained. One interesting implication is an estimate of effective  width  of the photosspheric turbulent layer that is probed by the detector. Conclusions are presented in section 5.

\section{ The 1-D power spectrum of  line-of-sight integrated data}

   We are interested in the power spectrum of  a quantity $n(x, D)$ (where x is  straight line in a lateral direction) which  is an integral along the line-of -sight  $z$ of a underlying physical quantity, $f(x, z)$. 
  \begin{equation}
  n(x, D) = \int_0 ^{D } f( x, z)dz$$ 
   \end{equation}   
with $D$ denoting the  effective width of the  region that contributes to  the observed $n(x, D)$. It is termed "effective" because it is tacitly assumed in equation (1) that the weight of $f(x, z)$ is   independent of $z$ . An ideal case is when the observed quantity is some radiation intensity emanating from  a  {\it optically thin} slab of width $D$. It is also "effective" because in real situations $D$ can differ for different $x$ values.  
   
 We are interested in the 1-dimensional power spectrum of $n(x, D)$ which depends also on the value of the depth $D$.
  
  \begin{equation}
  E(k_x, D) = \int_{-\infty}^{\infty} e^{i k_x x} C_n(x, D) dx 
  \end{equation}
   with  $C_n (x, D)$ being the  2-point autocorrelation of the fluctuating $n(x,D)$ ( mean of $n(x, D)=0$).
   
  \begin{eqnarray}
       C_n(x, D)= <n( x', D) n(x'+x, D) >=
      \\ \nonumber 
   \int_{0} ^{D }\int_{0} ^{D}< f(x', z')(f( x'+x,   z) >    dzdz' 
   \end{eqnarray}
   with the $<>$ brackets denoting ensemble average.
   Assuming isotropy and homogeneity the 2-point autocorrelation of $f(x, z)$
   is   $C_f(x, z-z') = <f(x', z')(f( x'+x,   z) >$.
   
   So that 
   
   \begin{eqnarray}
          C_n(x, D)= \int_{0} ^{D }\int_{0} ^{D}C_f(x, z-z') dz dz' = \\ \nonumber 
    \int_{0} ^D\int_0^D\left(\int_{-\infty}^{\infty} \int _{-\infty}^{\infty} P_2(k_x, k_z) e^{-i k_x x -i k_z( z-z')} dk_x dk_z\right)dz dz'
    \end{eqnarray}
    
    where $P_2(k_x. k_y)$ is the 2-dimensional power spectrum. Interchanging the integration order yields,
     
    \begin{eqnarray}
    C_n(x, D)= D^2 \int_{-\infty}^{\infty}  e^{-ik_x x}   \int _{-\infty}^{\infty} \left(\frac{\sin( k_z D/2)}{k_z D/2} \right)^2   P_2(k_x, k_z)\\
    \nonumber     dk_z dk_x
\end{eqnarray}     
Consider a 2-dimensional   power spectrum that is isotropic, homogeneous and in the form of a power law 
\begin{equation}
    P_2 ( k_x, k_y) = A \left(k_x^2 + k_z^2 \right)^{-(m+1)/2}
 \end{equation}    
where A is  a constant and  $m$ is the index of the 1-D power spectrum.  
 
 Combining equations~ (2, 5, 6)  one easily identifies
 \begin{eqnarray}
 E(k_x, D) = M\int _0^{\infty}\left(k_x^2 + k_z^2 \right)^{-(m+1)/2}\left(\frac{\sin( k_z D/2)}{k_z D/2}  \right)^2 dk_z
 \end{eqnarray}
 with M a constant. Using a variable  $ \eta= k_z D/2$,  equation~(7) takes the form 

 \begin{eqnarray} 
 E(k_x, D) = N\int _0^{\infty}\left((k_x D/2)^2 + \eta^2 \right)^{-(m+1)/2}\left(\frac{\sin(\eta}{\eta}  \right)^2 d\eta
  \end{eqnarray} 
   with $N$ a constant. 
 \section{analytical results}
 
 It follows from equation (8) that for   given values  of $m$ and $N$,  $E(k_x, D)$ is a function of the dimensionless product $k_x D/2$. We are interested in the shape of the power spectrum  and   not its absolute normalization so the actual value of $N$ does not matter. In what follows we show  {\it analytically} that the asymptotic shapes of the power spectrum  are

\begin{eqnarray}
E(k_x, D)\propto (k_x D/2) ^{-m};\  \ k_x D/2<<1 \\
 E(k_x, D)\propto (k_x D/2) ^{-m-1} ;\  \ k_x D/2>>1 \nonumber
  \end{eqnarray}  
 To that end note that the $(\sin(\eta)/\eta)^2$ term implies that the integration upper limit is $\sim \pi/2$. Therefore,

 \begin {eqnarray}  
 E(k_x, D) \sim N\int_0^{\pi/2} \left((k_x D/2)^2 + \eta^2 \right)^{-(m+1)/2} d\eta \\
=N (k_x D/2)^{-m}\int_0^{\pi/( k_x D)} \left((1+ \mu^2 \right)^{-(m+1)/2} d\mu\nonumber
  \end {eqnarray}
  with $ \mu= \eta/ (k_x D/2)$.

In the limit $k_x D/2<<1$ the upper integration limit tends to infinity and the integral is independent  of $k_x D/2$. In the limit  $k_x D/2>>1$ the integral is proportional to $(k_x D/2) ^{-1}$.

The transition occurs at a transition  wavenumber, $k_t$, such that  $k_t D/2 = O(1)$. 
The exact value of $k_t D/2$  depends on $m$ and is required in order to find  the value of $D$.
To get the exact value, for the present case,  $ m=1.2$,
  equation (8)  is solved numerically, using WolframMathematica.

 \begin{figure} 
                       \centerline{\includegraphics[width=\columnwidth]{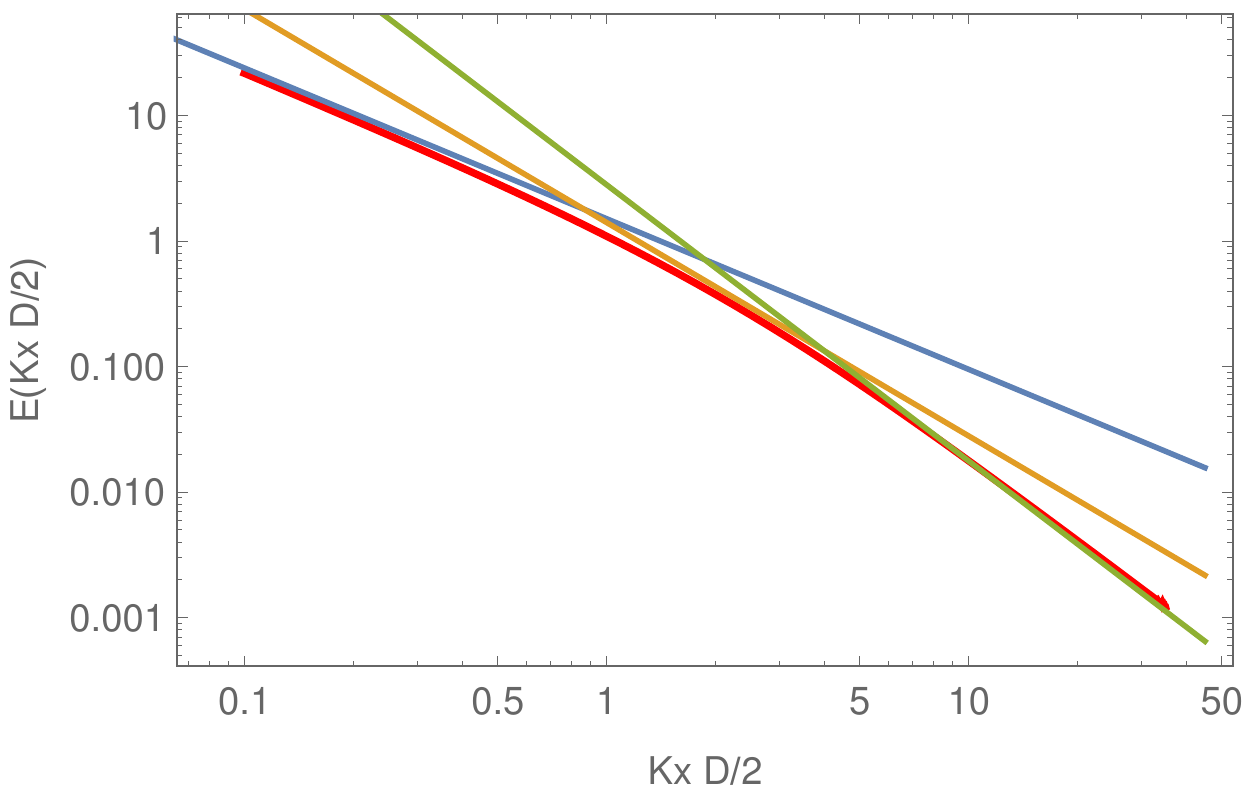}}
                       
 \caption{ { \bf  Thick  red   curve}: $E(k_x D/2)$  in arbitrary units as function of $k_x D/2$.   { \bf   blue line }: slope of -1.2. {\bf green line}: slope of -2.2.
 {\bf yellow line} : slope of -1.7.}
 \end{figure} 
 
   \section{numerical Results} 
 
 The   power spectrum $E(k_x D/2)$ for  $m=1.2$,  obtained by the numerical solution of equation (8),   is displayed in  Fig.~1.
       
       It is seen that for
 $k_x D/2<<1$ the power spectrum has a logarithmic slope of $-1.2$ while for   $k_x D/2>>1$ the logarithmic slope is $ -2.2$, as expected from the analytical analysis presented above.

  A line with a logarithmic slope of $-1.7$, the mean of the former two, which is tangent to 
 the curve defines the "point of transition". One should note that the transition is a smooth one, and not a sharp break at a point.
  
 From Fig. 1 it follows that the transition wavenumber, $k_t$, satisfies
\begin{equation}
 k_t D/2=  2.25 \pm 0.05 
\end{equation} 
  So a given
  {\it observational} value of $k_t$   determines $D$. 
  Expressing $k_t$by  $x_t=2\pi/k_t$ ,   the observational transition  spatial scale,  leads to
  \begin{equation}
  D=  (2.25 \pm 0.05)\frac{x_t}{\pi}
  \end{equation}
   \cite{Abramenko+Yurchyshyn2020}found  $x_t= 0.85\pm 0.05$~Mm.
 Thus,
\begin{equation}
 D=(0.57\pm 0.05){\rm \ Mm} \nonumber
\end{equation}

 \section{Concluding Remarks}
The paper by \cite{Abramenko+Yurchyshyn2020}provided  a very neat power spectrum of the photosphere magnetic field of the quiet Sun. The high spatial resolution of the NIRIS data enabled a power spectrum down to spatial scales as small as $0.3$~Mm. This power spectrum exhibits what appears as  two inertial ranges with logarithmic slopes of $-1.2$ and 
$-2.2$, for a spatial scale larger/smaller  than   $(0.85\pm 0.05)$~Mm, respectively.

 The logarithmic slope of -1.2 corresponds to a 3-~dimensional power spectrum with a logarithmic slope of    -3.2. This is significantly shallower that the -3.67 value corresponding to Kolmogorov turbulence.
The very interesting issue of the origin of such a power spectrum was not addressed here.

The present  paper demonstrates that two inertial ranges actually correspond  to a turbulence with a single inertial range with logarithmic slope equaling -1.2.
The two ranges are manifestation of the fact that the observed infra-red emission is an integral along the line of sight. Since the value of $D$   is within the span of the   lateral spatial scales probed, the transition shows up.

Interestingly, the value of the derived effective width is comparable to the width  of the photosphere  itself. This value of $D$ implied by the NIRIS power spectrum of  \cite{Abramenko+Yurchyshyn2020}
is an observational constraint to be faced with theoretical models of near IR lines in the quiet solar photosphere. 

In this context, a numerical model  by 
   \cite{Rueedi+98} yielded  that the FeI15648 Å  line is formed in the  lower part of the photosphere  and extends over a vertical span of about $0.1$~Mm. Their work  refers to a sun spot and not to the quiet sun, and    also assumes an absence of macro velocity turbulence. The situation here is different.

 \section*{Acknowledgments}       

 I thank Prof. Valentina Abramenko, the referee,  for very constructive remarks that helped
 to improve and  clarify  the paper.
This research has been supported by the Afeka College Research Fund.

\section*{data availability}
 The article presents  a theoretical research. It does apply the observational results of \cite{Abramenko+Yurchyshyn2020}
to the derived  theoretical results.

\label{lastpage}  
\end{document}